\documentclass[10pt,aps,prb,twocolumn,superscriptaddress,floatfix,nofootinbib]{revtex4-2}
\usepackage{amsmath}
\usepackage[english]{babel}
\usepackage{graphicx}
\usepackage[utf8]{inputenc}
\usepackage[colorlinks,linkcolor=blue,citecolor=blue,urlcolor=blue]{hyperref}

\begin{document}
\title{Evidence for valence-bond pairing in a one-dimensional two-orbital system}

\author{M. Mierzejewski}
\affiliation{Institute of Theoretical Physics, Faculty of Fundamental Problems of Technology, Wroc{\l}aw University of Science and Technology, 50-370 Wroc{\l}aw, Poland}

\author{E. Dagotto}
\affiliation{Department of Physics and Astronomy, University of Tennessee, Knoxville, Tennessee 37996, USA}
\affiliation{Materials Science and Technology Division, Oak Ridge National Laboratory, Oak Ridge, Tennessee 37831, USA}

\author{J. Herbrych}
\affiliation{Institute of Theoretical Physics, Faculty of Fundamental Problems of Technology, Wroc{\l}aw University of Science and Technology, 50-370 Wroc{\l}aw, Poland}

\begin{abstract}
Valence bond (VB) states as the formation mechanism of Cooper pairs, eventually leading to high-temperature superconductivity, remain a controversial topic. Although various VB-like states find variational relevance in the description of specific spin models and quantum spin liquids, in the realm of many-body fermionic Hamiltonians, the evidence for such states as ground states wave functions remains elusive, challenging the valence-bond pairing mechanism. Here, we present evidence of a VB ground state with pairing tendencies, particularly at finite doping. We achieved this for the generic two-orbital Hubbard model in low dimension, where the VB states can be associated with the presence of the topological order manifested by edge states.
\end{abstract}

\maketitle

{\it Introduction.}--In 1987, just one year after the discovery of high-temperature superconductivity (high-$T_\mathrm{c}$ SC) in cuprates \cite{BednorzMuller1986}, Philip W. Anderson proposed \cite{Anderson1987} his famous resonating valence bond (RVB) state as the ground state wavefunction to describes the properties of such compounds. In essence, the RVB represents a quantum liquid of valence bonds, i.e., a collection of spin singlets, distributed over the lattice in a way that preserves its spatial symmetries. Note that the singlets are not necessarily nearest neighbours and can span over a few lattice sites, though their amplitude is expected to decay exponentially with distance. Such a state can, in principle, describe the Mott insulators' spin arrangement for the half electronic filing \cite{Liang1988} and, more importantly, allow for mobile Cooper pairs under hole or electron doping (with each pair of holes/electrons "replacing" one of the singlets) \cite{Baskaran1987,Anderson1987-2}.

\begin{figure*}[!ht]
\includegraphics[width=0.95\textwidth]{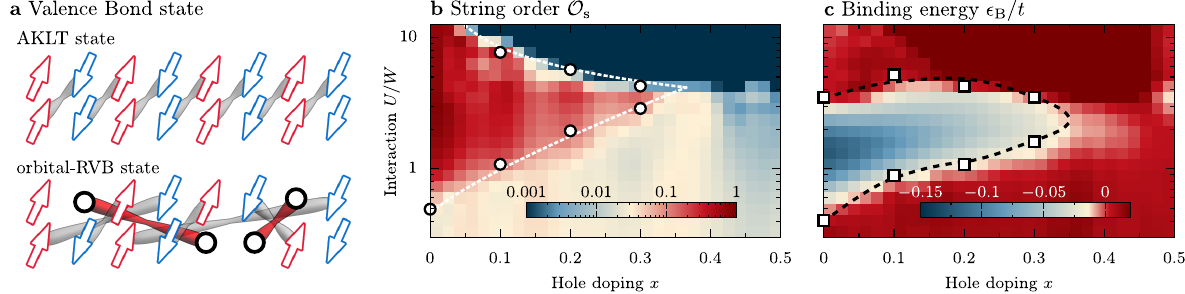}
\caption{{\bf a} Sketch of the AKLT state (top) and the orbital-RVB state of the doped one-dimensional two-orbital Hubbard model (lower). Paired holes are presented as circles. {\bf b} Interaction $U$ -- doping $x$ phase diagram of the string order parameter $\mathcal{O}_\mathrm{s}$ evaluated in the bulk (at distance $\ell=L/2$) of the $L=60$ sites system. The points depict values at which extrapolated to $L\to\infty$ spin gap $\Delta_\mathrm{S}$ opens or closes. {\bf c} Phase diagram for binding energy \mbox{$\epsilon_\mathrm{B}=E_\mathrm{gs}(N)-2E_\mathrm{gs}(N-1)+E_\mathrm{gs}(N-2)$}, where $E_\mathrm{gs}(N)$ is the energy of the fermionic system with $N$ electrons ($L=60$). Points depict values at which the extrapolated to $L\to\infty$ binding energy crosses the zero value. Lines are a guide to the eye.}
\label{fig1}
\end{figure*}

Anderson's idea sparked enormous interest in the antiferromagnetic (AFM) Heisenberg models, which properly describe the main experimental findings of undoped cuprates \cite{Kageyama1999,Piazza2015} and parent compounds of some iron-based superconductors \cite{Glasbrenner2015}. However, RVB and even more generic valence bond solids (VBS, which break some of the lattice symmetry) are rare as the ground states wavefunctions of the many-body systems. For the spin models, notable examples of such states exist: (i)~it is by now established that quantum spin liquids \cite{Savary2017} realized in geometrically frustrated magnets \cite{Balents2010,Wang2015} can be described by the RVB wave function. (ii)~The ground states of the one-dimensional (1D) $S=1/2$ Heisenberg model with nearest- and next-nearest-neighbour interaction (the so-called Majumdar-Ghosh model) \cite{Majumdar1969-1,Majumdar1969-2} or the two-dimensional (2D) Shastry-Sutherland  model \cite{Shastry1981} are exact VBS states. (iii)~Finally, it was shown \cite{Affleck1987,Affleck1988} that the $S=1$ Heisenberg chain with biquadratic interactions can be thought of as a collection of coupled $S=1/2$-like singlets (Fig.~\ref{fig1}a). The latter is encapsulated in the Affleck-Kennedy-Lieb-Tasaki state (AKLT state), which hosts the famous topological Haldane edge states. It is important to note that the AKLT state is a perfect VBS. At the same time, the plain $S=1$ Heisenberg model in 1D resembles a gapped RVB state with exponentially decaying correlations \cite{Affleck1988-2}.

In the context of fermionic Hamiltonians, evidence of a VB-like state as the ground state of the many-body system is still lacking. Although such states are anticipated to capture many properties of quantum paramagnets, the challenge is demonstrating that the ground state has a form of VB liquid. Here, we report the evidence that the ground state of the generic two-orbital Hubbard model in low-dimension realizes a VB-like state. The latter maintains the topological properties of the AKLT state and, in addition, becomes superconducting at finite doping. {\it Namely, we show that the topology and pairing are intertwined in the doped fermionic Haldane chain.} Although limited to 1D considerations, our density-matrix renormalization group (DMRG) calculations reveal most of the properties expected of the high-$T_\mathrm{c}$ phase diagram, i.e., (i)~a large region of finite binding energy in the interaction-doping phase diagram; (ii)~long-range of pair-pair correlations; and (iii)~pair-density-wave (PDW). We also show that (iv)~all of these phenomena are induced by the VB ground state akin to the AKLT state of $S=1$ Heisenberg chains. Since the latter exhibits topological properties, the presence of VB-type states can be easily identified via the presence of the topological order.

{\it Orbital-RVB.}--Our investigation is based on the two-orbital ($\gamma=0,1$) Hubbard-Kanamori model on the 1D lattice:
\begin{eqnarray}
H&=& t\sum_{\gamma\gamma^\prime\ell\sigma}
\left(c^{\dagger}_{\gamma\ell\sigma}c^{\phantom{\dagger}}_{\gamma^\prime\ell+1\sigma}+\mathrm{H.c.}\right)\nonumber\\
&+&
U\sum_{\gamma\ell}n_{\gamma\ell\uparrow}n_{\gamma\ell\downarrow}
+U^\prime \sum_{\ell} n_{0\ell} n_{1\ell}\nonumber\\
&-& 2J_{\mathrm{H}} \sum_{\ell} \mathbf{S}_{0\ell} \cdot \mathbf{S}_{1\ell}
+J_{\mathrm{H}} \sum_{\ell} \left(P^{\dagger}_{0\ell}P^{\phantom{\dagger}}_{1\ell}+\mathrm{H.c.}\right)\,,
\label{hamhub}
\end{eqnarray}
with $P^\dagger_{\gamma\ell}=c^\dagger_{\gamma\ell\uparrow}c^\dagger_{\gamma\ell\downarrow}$. In the following, we will consider its most generic version, with band degeneracy. The first term describes the system's kinetic energy (with kinetic energy span given by $W=4t$). The second term describes intra- ($U$) and inter-orbital ($U^\prime$) on-site electron repulsion. The last term originates in multi-orbital physics: $J_\mathrm{H}$ accounts for the ferromagnetic Hund coupling between spins $\mathbf{S}_{\gamma\ell}$ at different orbitals, maximizing the total on-site spin. The above model preserves SU(2) symmetry (provided that $U^\prime=U-5/2J_\mathrm{H}$ \cite{Georges2013}), and we will consider $J_\mathrm{H}/U=1/4$ in the $S^z_\mathrm{tot}=0$ magnetization sector for various hole doping levels $x=1-\overline{n}$, where $\overline{n}=N/2L$ is electron density (with $N$ as the number of electrons in a $L$ site system). The quasi-1D (ladders) and 2D versions of the above model are extensively used in the context of various correlated superconductors like iron pnictides, chalcogenides, ruthenates, iridates, as well as heavy-fermion materials. In the following, we will present results obtained with the help of the DMRG method on the open boundary system (see Supplemental Material \cite{supp} and references \cite{White1992,Schollwock2005,White2005} therein for details).

At half-filling $x=0$ and in the limit of large interaction strength $U\gg W$, i.e., in the region where double occupancies are not present and the average on-site magnetic moments are well developed \mbox{$\mathbf{S}^2=S(S+1)\simeq2$}, the low energy physics of the two-orbital Hubbard model can be described by the $S=1$ AFM Heisenberg Hamiltonian \cite{Jazdzewska2023}. The ground state of the latter can be pictorially expressed (Fig.~\ref{fig1}a) by on-site triplets of $S=1/2$-like objects, i.e., 
\begin{eqnarray*}
|1_i\rangle=|\uparrow_{0i}\uparrow_{1i}\rangle\,,\quad|-1_i\rangle=|\downarrow_{0i}\downarrow_{1i}\rangle\,,\\
|0_i\rangle=\frac{1}{\sqrt{2}}\left(|\uparrow_{0i}\downarrow_{1i}\rangle+|\downarrow_{0i}\uparrow_{1i}\rangle\right)\,,
\end{eqnarray*}
which are coupled in a valence bond way between sites, $(1/\sqrt{2})(|\uparrow_{0i}\downarrow_{1i+1}\rangle-|\downarrow_{0i}\uparrow_{1i+1}\rangle)$. Here $|1,0,-1\rangle$ represent $S=1$ states at site $i$, while $|\sigma_{\gamma i}\sigma_{\gamma^\prime j}\rangle$ with $\sigma_i=\uparrow,\downarrow$ can be thought as a spin configuration of two electrons at different orbitals $\gamma\,,\gamma^\prime$ in the context of the two-orbital Hubbard model. The above VBS (AKLT state) is not an exact ground state of the isotropic $S=1$ AFM model. Still, it can be adiabatically connected to it without closing the spin gap and preserving its unique properties, i.e., the presence of the topologically protected Haldane edge states (unpaired by valence bonds $S=1/2$ states at the boundary of the open system). It was recently shown \cite{Jazdzewska2023} that such a description of the two-orbital Hubbard model in 1D is valid even for a relatively small value of interaction, $U\simeq W/2$, in the region where magnetic moments are not fully developed $\mathbf{S}^2\ll2$ and charge fluctuations are still present. Such behaviour can be monitored with the help of the string order correlation function: 
\begin{equation}
\mathcal{O}_\mathrm{s}(\ell)=-\left\langle \mathrm{gs}\right|S^z_m \exp\left(i\pi\sum_{n=m+1}^{m+\ell-1} S^z_{n}\right) S^z_{m+\ell}\left|\mathrm{gs}\right\rangle\,,
\label{string}
\end{equation}
with $|\mathrm{gs}\rangle$ as the ground state wavefunction and $S^z_m=S^z_{0m}+S^z_{1m}$ as the total spin at site $m$, which serves as the order parameter of the AKLT state in the $\ell\to\infty$ limit (i.e., breaking of the discrete $Z_2\times Z_2$ hidden symmetry).

\begin{figure}[!ht]
\includegraphics[width=1.0\columnwidth]{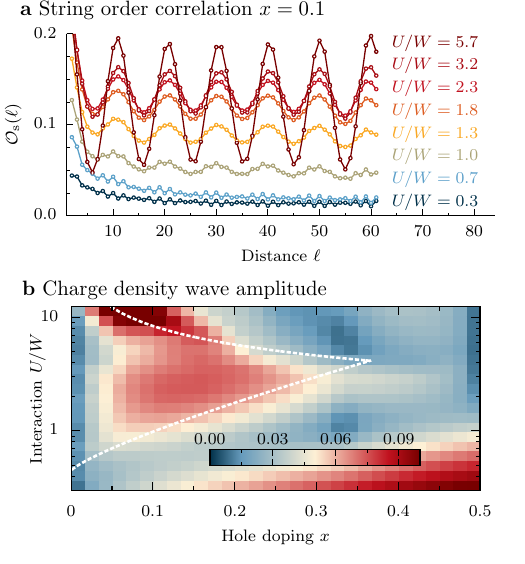}
\caption{{\bf a} Distance dependence of string order correlations function $\mathcal{O}_\mathrm{s}(\ell)$ for various interaction strength $U/W$ and $x=0.1$ hole doped system ($L=80$ data). {\bf b} Amplitude of the charge density oscillations ($L=60$). Lines represent the same guide to the eye as in Fig.~\ref{fig1}.}
\label{fig2}
\end{figure}

Introduction of the holes into the AKLT state is a nontrivial task. In principle, two scenarios are possible in the atomic limit: the formation of "rigid" on-site holes $|h_{0i}h_{1i}\rangle$ in the AFM $S=1$ background or the pair of holes transforming two of the $S=1$ on-site triplets into two $S=1/2$ objects, e.g.,
\begin{eqnarray*}
|\uparrow_{0i}\downarrow_{1i}\rangle-|\downarrow_{0i}\uparrow_{1i}\rangle
\quad\to\quad
|\uparrow_{0i}h_{1i}\rangle-|h_{0i}\uparrow_{1i}\rangle.
\end{eqnarray*}
In the atomic limit $t=0$, the former scenario is favoured by the Hund exchange $J_\mathrm{H}$ since it's maximizing the average magnetic moments. On the other hand, the latter case - with one hole per site - is preferred by the inter-orbital repulsion $U^\prime$ term in the Hamiltonian. To resolve this state in the many-body system $t\ne0$, we monitor the behaviour of the ground state with doping $x$ and interaction $U$ by evaluating the string order parameter $\mathcal{O}_s(\ell)$. Detailed distance dependence is presented in Fig.~\ref{fig2}a, while Fig.~\ref{fig1}b depicts the phase diagram obtained from the bulk value of $\mathcal{O}_\mathrm{s}(L/2)$ (see also Supplemental Material \cite{supp} for additional results). For dopings $x\lesssim 0.35$, one can observe a region where the string order parameter is finite, excluding two holes at the same site (but different orbitals $|h_{i0}h_{i1}\rangle$), which would break the AFM chain into two parts. More importantly, our results are consistent with a one hole per site scenario, i.e., with (two) holes replacing one of the valence bonds. For a few holes away from half-filling, $x=0$, such a state was coined the orbital-RVB \cite{Patel2017,Patel2019,Laurell2024}. Here, we extend this definition to rather large doping levels, $x\simeq0.35$, and show that its properties are consistent with the valence-bond pairing mechanism.

\begin{figure}[!ht]
\includegraphics[width=1.0\columnwidth]{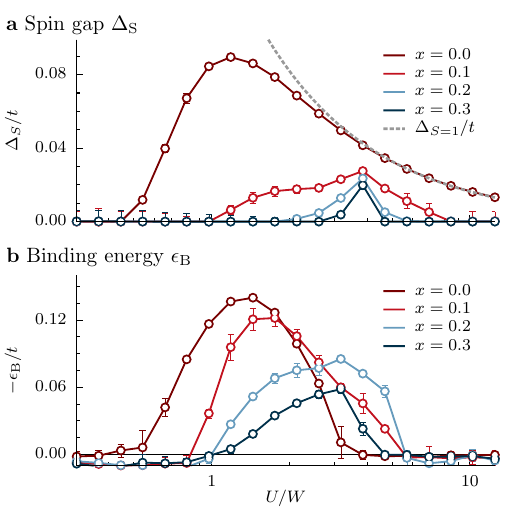}
\caption{Extrapolated to $L\to\infty$ {\bf a} spin gap $\Delta_\mathrm{S}=E_\mathrm{gs}(S^z_\mathrm{tot}=2)-E_\mathrm{gs}(S^z_\mathrm{tot}=0)$ and {\bf b} binding energy $-\epsilon_\mathrm{B}$. Dashed line in {\bf a} indicates Haldane gap $\Delta_\mathrm{S=1}=0.41J$ with $J=2t^2/(U+J_\mathrm{H})$.}
\label{fig3}
\end{figure}

For $x=0$ (half-filling), one can observe (Fig.~\ref{fig1}b) a finite string-order parameter for all $U/W\gtrsim 0.5$ (with $U\to\infty$ limit given by the $S=1$ Heisenberg model), while for $x\ne0$ this is true only in a finite region of interaction strength $U$. One can understand the lower (at small $U$) topological phase transition (from trivial paramagnetic to topological orbital-RVB state) as an effect of interaction $U$ strengthening the magnetic moments $\mathbf{S}$ and decreasing the charge fluctuations. On the other hand, the upper (at large $U$) transition is associated with the change in the spin-spin correlations from AFM (incommensurate at finite $x$) to ferromagnetic (FM) ordering due to double-exchange like physics in the large Hund limit for $x\ne0$ \cite{Momoi1998,Held2000}. In Supplementary Material (see also references \cite{Zener1951,Zener1951-2,Assaad1991,Troyer1993,Lin1997,Riera1997,Haule2009,Luo2010,Ferber2012} therein), we also present the analysis of the correlations between the system's edges, confirming the presence of topologically protected edge states, and the spin-spin magnetic structure factor analysis confirming the AFM to FM transition at large $U$. 

Since the orbital-RVB state shares its topological properties with the "standard" AKLT state, one expects that finite string order $\mathcal{O}_s\ne0$ is accompanied by the presence of the reminiscence of the Haldane gap [\mbox{$\Delta_\mathrm{S=1}\simeq0.41J$} for the $x=0$ AKLT state, with spin exchange \mbox{$J=2t^2/(U+J_\mathrm{H})$} \cite{Jazdzewska2023}]. In Fig.~\ref{fig3}a, we present the interaction $U$ dependence of the (extrapolated to thermodynamic limit \cite{supp}) spin gap $\Delta_\mathrm{S}$. Indeed, our results indicate that $\Delta_\mathrm{S}$ remains open in the same region where the string order is finite, also for nonzero doping $x\ne0$ (see points presented in Fig.~\ref{fig1}b, which depict opening and closing of the spin gap).

\begin{figure*}[!ht]
\includegraphics[width=1.0\textwidth]{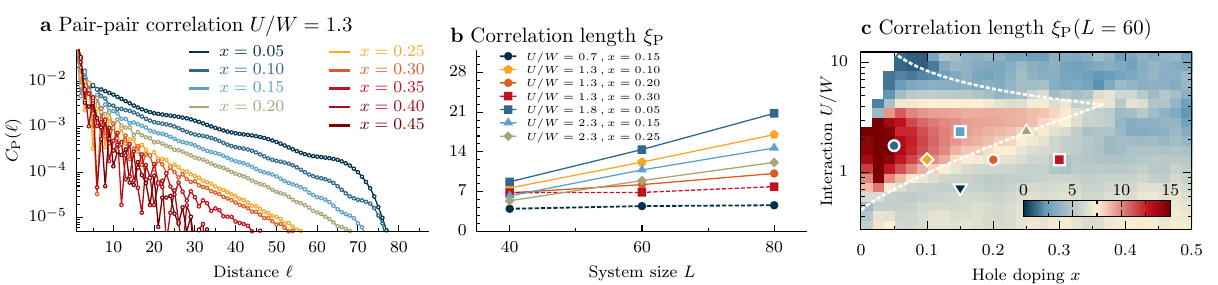}
\caption{{\bf a} Doping $x$ dependence of the pair-pair correlation function $C_\mathrm{P}(\ell)$ for $U/W=1.3$ ($L=80$ data). {\bf b} Finite-size scaling ($L=40,60,80$) of pair-pair correlation length $\xi_\mathrm{P}$ obtained from $C_\mathrm{P}(\ell)\propto \exp(-\ell/\xi_\mathrm{P})$ fits (see Supplementary Material for more data \cite{supp}). {\bf c} Phase diagram of the correlation length $\xi_\mathrm{P}$ obtained from the fits to the $L=60$ sites system. Points depict parameters presented in panel {\bf b}. Lines represent the same lines represent guides to the eye as in Fig.~\ref{fig1}.}
\label{fig4}
\end{figure*}

Finally, the spatial dependence of the string order parameter $\mathcal{O}_\mathrm{s}(\ell)$ indicates the presence of pronounced oscillations for $x\ne0$. We associate them with the presence of Friedel oscillations in the charge sector (due to the open boundary system), visible also in the site-resolved electron density $n_\ell=n_{0\ell}+n_{1\ell}$ (see the Supplementary Material \cite{supp}). In Fig.~\ref{fig2}b, we present the spatial standard deviation of the density $\sigma_n=\sqrt{(1/L)\sum_\ell n_\ell^2-\overline{n}^2}$, related to the amplitude $A$ of the cosine-wave, $\sigma_n\simeq A/\sqrt{2}$. In the topologically trivial region, $U\lesssim W$ and $x\gtrsim0.2$ we find the "standard" $2k_\mathrm{F}=\pi\overline{n}=\pi(1-x)\propto \pi x$ charge-density-wave oscillations of a weakly interacting system. Interestingly, in the region where we find finite $\mathcal{O}_\mathrm{s}\ne0$ ($U\sim2W$, $x\lesssim 0.35$), our results indicate that the charge oscillates with $2\pi x$ wave-vector. The latter is consistent with a pair-density wave (PDW) with $4k_\mathrm{F}$ oscillations \cite{Berg2010,Zhang2022,Liu2023} (leading to two doped holes per one minimum in oscillation). This indicated that the PDW accompanies the doped orbital-RVB and is most pronounced at doping $x\simeq 0.15$. 

{\it Binding energy.}--Our results show that the orbital-RVB state remains robust even for large doping $x\simeq0.3$ at $U\simeq 4W$. In Fig.~\ref{fig1}c, we contrast this behaviour with the binding energy $\epsilon_\mathrm{B}$, where a negative value signals the presence of bound pairs of holes (or electrons for $x<0$, due to the particle-hole symmetry of the considered model). The significant overlap in the interaction-doping phase diagram between the finite string order, $\mathcal{O}_\mathrm{s}\ne0$, and negative binding energy, $\epsilon_\mathrm{B}<0$, indicates that the VB structure of the ground state and bounded pairs coexist. Consequently, the latter are correlated due to the former (see Fig.~\ref{fig1}a). The above behaviour is in accord with the scenario envisioned by Anderson's proposal: upon doping, the system minimises its energy by breaking a minimal amount of coupled singlets (valence bonds) \cite{Dagotto1992}. Interestingly, the VB-induced pairing (as measured by negative binding energy) is found for the electron-electron interaction strengths $U$, which are "just below" the large interaction expansion, i.e., below the $J(U)\propto1/U$ energy scale, which gives the proper description of the spin excitations.

It is important to note that the presence of the bounded pairs does not necessarily imply superconducting tendencies. The latter requires non-vanishing long-range correlations in the thermodynamic limit $L\to\infty$. One can monitor this with Cooper pair susceptibility
\begin{equation}
C_\mathrm{P}(\ell)=\frac{1}{L-\ell}\sum_{i,\gamma\ne\gamma^\prime}
\left\langle
\Delta^{\dagger}_{i\gamma\gamma^\prime}\Delta^{\phantom{\dagger}}_{i+\ell\gamma\gamma^\prime}
\right\rangle\,,
\end{equation}
where $\Delta^\dagger_{i\gamma\gamma^\prime}$ represent singlet pairs between nearest-neighbour sites at different orbitals, $\Delta^\dagger_{i\gamma\gamma^\prime}=c^\dagger_{\gamma i\uparrow}c^\dagger_{\gamma^\prime i+1\downarrow}-c^\dagger_{\gamma i\downarrow}c^\dagger_{\gamma^\prime i+1\uparrow}$. Such a pairing is consistent with doping of the orbital-RVB state described above and with earlier numerical investigations \cite{Patel2017}. Our results indicate two distinct behaviors; see Fig.~\ref{fig4}(a,b) and the Supplemental Material \cite{supp}. For the trivial (non-VB) state (e.g., $x\gtrsim0.25$ for $U=1.3$), we observe fast exponential decay of pair-pair correlation $C_\mathrm{P}(\ell)\propto \exp(-\ell/\xi_\mathrm{P})$, with size independent correlation length $\xi_\mathrm{P}$ (see Fig.~\ref{fig4}b). On the other hand, for the orbital-RVB states we find that the correlation length increases with system size, $\xi_\mathrm{P}(L)\propto L^\alpha$ with $\alpha\simeq 1$. The analysis of the $\xi_\mathrm{P}$ value for a finite-size system (Fig.~\ref{fig4}c) confirms that the correlation length in this region becomes large $\xi_\mathrm{P}(L=60)\sim 20$. This implies that even in the thermodynamic limit $L\to\infty$, the pairs in our VB state are correlated at distances of the order of the system size, again confirming Anderson's RVB pairing scenario.

{\it Conclusions.}--Since our results are obtained on the one-dimensional two-orbital model with total $S=1$ magnetic moments in the $U\gg W$ limit, the relevance of our results to cuprates is unknown (fundamentally considered as a single-orbital 2D system with $S=1/2$). However, it is well established that the multi-orbital nature of the Fermi surface plays a crucial role in the properties of iron-based superconductors. Consequently, the studied model is relevant for the latter. Consider the flagship iron-based superconductor Fe(Se,Te) \cite{Wang2016,Kreisel2020}. The magnetism of this compound is believed \cite{Wysocki2011,Wang2015,Glasbrenner2015} to be described by the frustrated $J_1$-$J_2$ $S=1$ Heisenberg model in 2D. Interestingly, for the relevant $J_2/J_1\sim0.5$ values, the system ground-state can be described \cite{Wang2015,Watson2016,Niesen2017} by spontaneously forming $S=1$ AKLT-like chains. Furthermore, it is known \cite{Pollmann2012} that odd-integer spin chains have topologically protected AKLT states. The latter can also be realized as frustrated multi-odd-leg $S=1$ tubes \cite{Charrier2010}. In essence, with caveats that our minimal model does not consider, e.g., the nematic phase transition present upon doping, our results are "just" doping of such systems.

Our findings reveal an astonishing robustness of the Haldane physics of the $S=1$ AFM chain upon hole doping (topologically nontrivial orbital-RVB state) and its importance for the pairing correlations even at significantly large doping levels ($x\simeq0.35$). This is unexpected since such a phase is fragile for pure spin models \cite{Anfuso2007,Moudgalya2015}. Furthermore, our results encapsulate the main features expected in the superconductor's phase diagram at zero temperature: doping the quantum paramagnetic system leads to a finite region in which pair-pair correlations are present. This is especially appealing for $U\sim 4W$ ($ U\sim 16t$), for which we see the above behaviour for $0.1\lesssim x\lesssim 0.3$. Furthermore, for $U\sim2W$ and $0.2<x<0.35$, the pair-pair correlation decays relatively fast, although the binding energy is negative $\epsilon_\mathrm{B}<0$. One can associate this with zero spin-gap $\Delta_\mathrm{S}$ and pre-formed pairs in the vicinity of a trivial-topological transition (which becomes coherent at larger $U$, when $\Delta_\mathrm{S}\ne0$, i.e., in the topologically nontrivial region). However, our finite-size data cannot exclude a small but finite $\Delta_\mathrm{S}$ already in this region. Finally, it is worth noting that the orbital-RVB state has a topological origin and is protected from perturbations that don't close the spin gap. Consequently, the specific values of the Hamiltonian parameters like Hund exchange $J_\mathrm{H}$ or details of the kinetic energy and hybridization between orbitals will not change the overall properties of the system, as was previously shown for half-filling $x=0$ \cite{Jazdzewska2023}.

\begin{acknowledgments}
J.~Herbrych acknowledges grant support by the National Science Centre (NCN), Poland, via Sonata BIS project no. 2023/50/E/ST3/00033. The US Department of Energy, Office of Science, Basic Energy Sciences, Materials Sciences and Engineering Division supported E. Dagotto. Part of the calculations have been carried out using resources provided by the Wroclaw Centre for Networking and Supercomputing (\url{http://wcss.pl}).\\
The data that support the findings of this article are openly available \cite{opendata}.
\end{acknowledgments}

\bibliography{haldanestability}

\clearpage
\appendix
\setcounter{page}{1}
\setcounter{figure}{0}
\setcounter{equation}{0}
\newcommand{\rom}[1]{\uppercase\expandafter{\romannumeral #1\relax}}
\renewcommand{\refname}{Supplemental References}
\renewcommand{\figurename}{Supplemental Figure}
\renewcommand{\thefigure}{S\arabic{figure}}
\renewcommand{\theHfigure}{S\arabic{figure}} 
\renewcommand{\citenumfont}[1]{#1}
\renewcommand{\bibnumfmt}[1]{[#1]}
\renewcommand{\thepage}{S\arabic{page}}
\renewcommand{\theequation}{S\arabic{equation}}
\renewcommand{\theHequation}{S\arabic{figure}} 
\renewcommand{\thetable}{S\Roman{equation}}
\renewcommand{\theHtable}{S\arabic{equation}} 
\onecolumngrid

\begin{center}
{\bf \uppercase{Supplementary Information} for}\\
\vspace{3pt}
{\bf \large \makeatletter\@title\makeatother}\\
\vspace{10pt}
by M. Mierzejewski, E. Dagotto, and J. Herbrych
\end{center}
\vspace{10pt}

\section{DMRG method}
\begin{figure*}[!htb]
\includegraphics[width=1.0\textwidth]{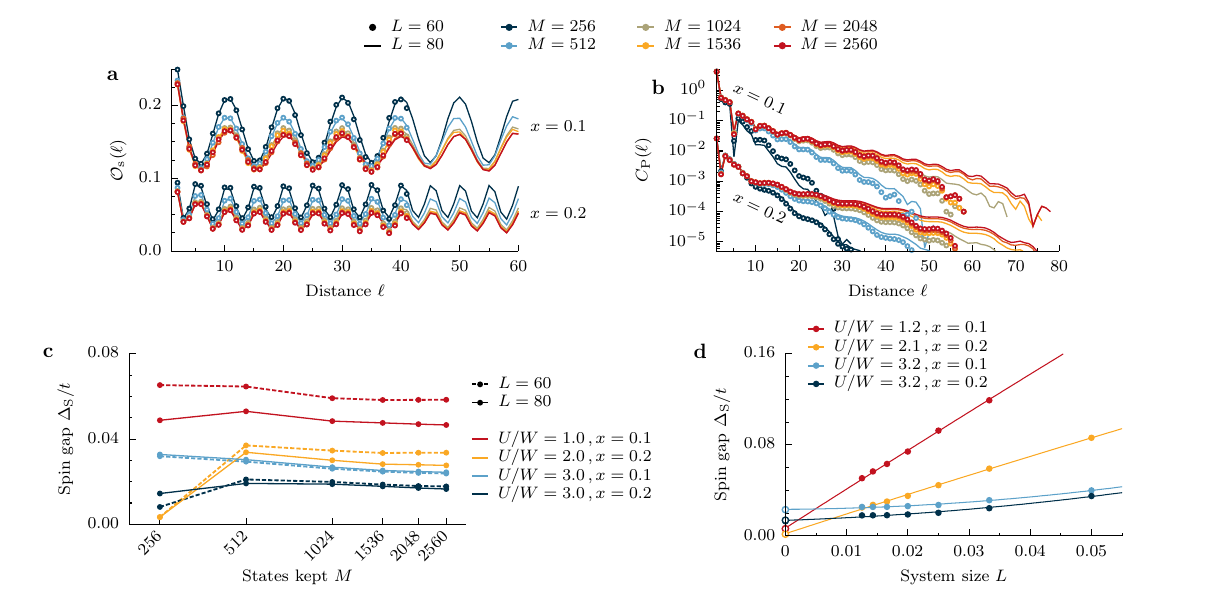}
\caption{{\bf Accuracy analysis.} {\bf a,b} System size $L=60\,,80$ and number of states kept in DMRG algorithm $M=256\,,512\,,1024\,,1536\,,2048\,,2560$ dependence of {\bf a} the string order parameter $\mathcal{O}_\mathrm{s}(\ell)$ and {\bf b} pair-pair correlation function $C_\mathrm{P}(\ell)$ for $U/W=3$ and $x=0.1\,,0.2$. {\bf c,d} Similar accuracy analysis of the spin gap $\Delta_{S}$ for selected values of $(U\,,n)$ parameters. Panel {\bf c} depict $M$ dependence, while {\bf d} shows $1/L$ scaling.}
\label{figS0}
\end{figure*}

All results presented in this work were obtained with the help of the zero-temperature density matrix renormalization group (DMRG) method\footnote[20]{S. R. White, "{Density matrix formulation for quantum renormalization groups}," \href{https://link.aps.org/doi/10.1103/PhysRevLett.69.2863}{Phys. Rev. Lett. {\bf 69}, 2863 (1992)}.}\footnote[21]{U. Schollw\"ock, "{The density-matrix renormalization group}," \href{https://link.aps.org/doi/10.1103/RevModPhys.77.259}{Rev. Mod. Phys. {\bf 77}, 259 (2005)}.} within the single-site algorithm\footnote[22]{S. R. White, "{Density matrix renormalization group algorithms with a single center site}," \href{https://link.aps.org/doi/10.1103/PhysRevB.72.180403}{Phys. Rev. B {\bf 72}, 180403 (2005)}.}. We performed at least $15$ sweeps, and used $A=0.001$ vector-offset in the single-site DMRG approach. The accuracy analysis, presented in Fig.~\ref{figS0}, indicates that $M=2048$ states kept is sufficient for all quantities discussed in our work. Such parameters allow us to accurately simulate system sizes up to $L\leq 80$ sites of the two-orbital Hubbard model.

\newpage
\section{Correlation functions}

\begin{figure*}[!htb]
\includegraphics[width=0.93\textwidth]{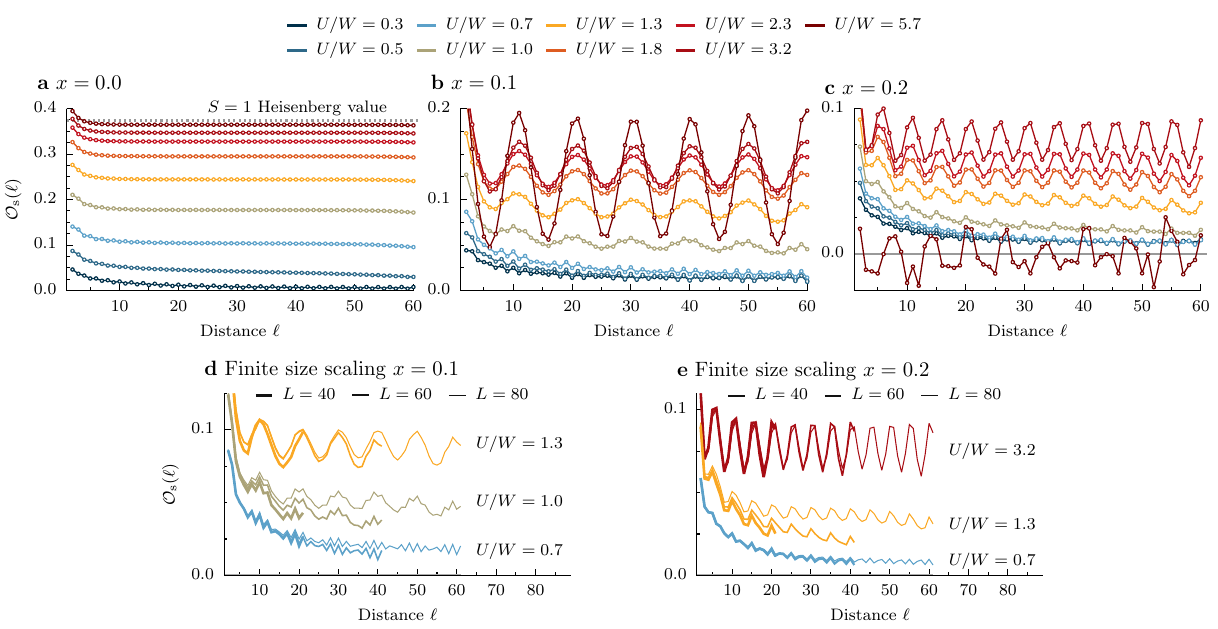}
\caption{{\bf String order correlations $\mathcal{O}_\mathrm{s}$.} {\bf a} Distance dependence of $\mathcal{O}_\mathrm{s}(\ell)$ for various interaction strength $U/W$ and half-filled system $x=0.0$. Change in the behaviour of $\mathcal{O}_\mathrm{s}$ is evident for $U/W\simeq 0.5$, i.e., for the predicted \cite{Jazdzewska2023} transition to the topological Haldane phase in which the string order is not decaying for $\ell\to\infty$. {\bf b,c} Similar results for hole doped system: $x=0.1$ and $x=0.2$, respectively. All results evaluated for $L=80$ site system. {\bf d,e} Finite-size scaling ($L=40,60,80$) for doped systems, {\bf d} $x=0.1$ and {\bf e} $x=0.2$, close to the transition.}
\label{figS3}
\end{figure*}

\begin{figure*}[!htb]
\includegraphics[width=0.93\textwidth]{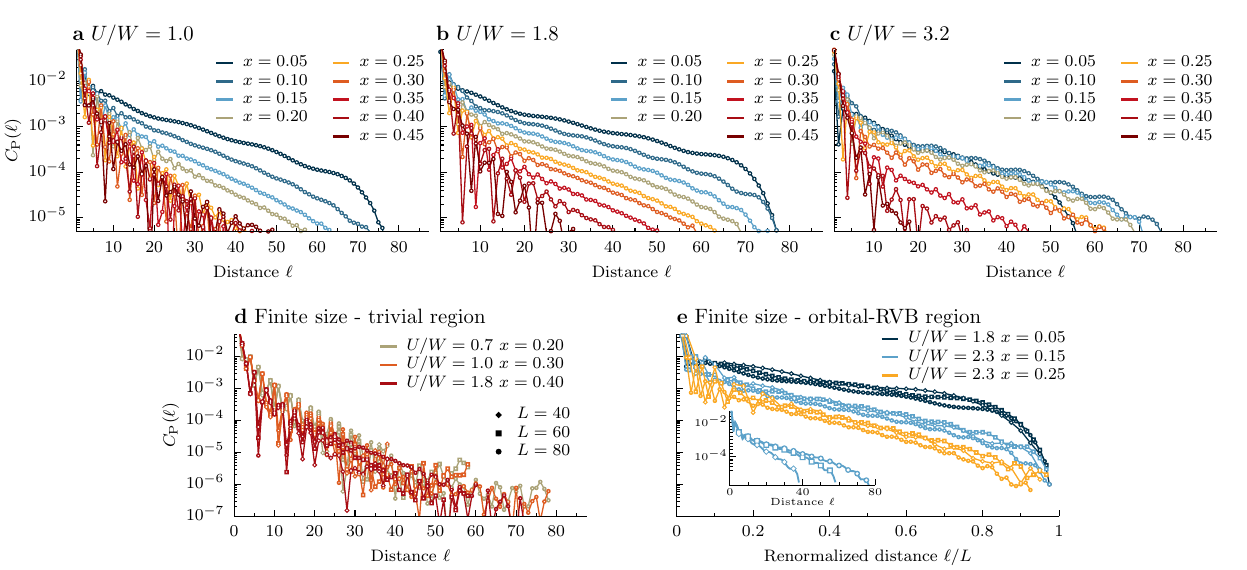}
\caption{{\bf Pair-pair correlation.} {\bf a,b,c} Doping $x$ dependence of the pair-pair correlation function $C_\mathrm{P}(\ell)$ for $U/W=1.0, 1.8,3.2$, respectively ($L=80$ data). {\bf d,e} Finite-size scaling $L=40,60,80$ of $C_\mathrm{P}(\ell)$ in {\bf d} trival (non-VBS) region and {\bf e} orbital-RVB region. Inset in the latter depicts results without renormalization of distance $\ell$, clearly showing system size dependent correlation length $\xi_\mathrm{P}(L)$.}
\label{figS4}
\end{figure*}

In Fig.~\ref{figS3} and Fig.~\ref{figS4} we present additional results for the string order correlation function $\mathcal{O}_\mathrm{s}(\ell)$ (Fig.~\ref{fig2} of the main text) and the pair-pair correlation function $C_\mathrm{P}(\ell)$ (Fig.~\ref{fig4} of the main text), respectively

\newpage
\section{Static spin correlations}

\begin{figure}[!htb]
\includegraphics[width=1.0\textwidth]{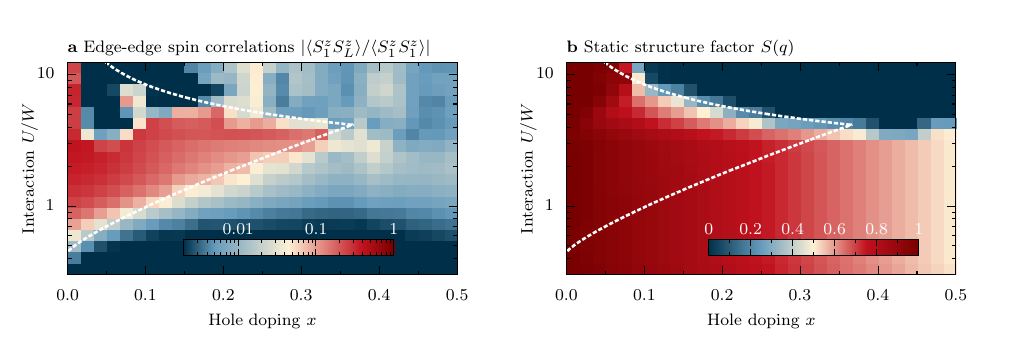}
\caption{{\bf Spin correlations.} Interaction $U$-doping $x$ phase diagram of {\bf a} edge-edge spin correlations $|\langle S^z_1S^z_L\rangle/\langle S^z_1S^z_1\rangle|$, and {\bf b} position of the maximum $q_\mathrm{max}/\pi$ of the static structure factor $S(q)$.}
\label{figS1}
\end{figure}

Figure~\ref{figS1} depicts additional results in the interaction $U$ - doping $x$ plane, i.e., (a) the edge-edge spin correlations and (b) the analysis of the static structure factor.

(a) Fig.~\ref{figS1}\textbf{a}. The behaviour of spin edge-edge correlations $|\langle S^z_1S^z_L\rangle/\langle S^z_1S^z_1\rangle|$ (with $S^z_\ell=S^z_{0\ell}+S^z_{1\ell}$) is consistent with the results presented in Fig.~\ref{fig1}{\bf b} of the main text. For the $(U,x)$ values for which the orbital-RVB state is found, the spin-spin correlations across the system (between the edges) are finite. The latter indicated the presence of the Haldane edge state, inherited from the $x=0$ and $U,J_\mathrm{H}\gg t$ limit of the $S=1$ Heisenberg model.

Special attention is needed in the small but nonzero doping region, $x\to 0$, and large interaction strength, $U/W>3$. We find phase separation between hole-undoped antiferromagnetic and hole-rich ferromagnetic regions. Such phase is expected in the systems with strong AFM correlations\footnote[32]{F. F. Assaad and D. W\"urtz, "{Charge and spin structures in the one-dimensional $t$-$J$ model}," \href{https://doi.org/10.1103/PhysRevB.44.2681}{Phys. Rev. B {\bf 44}, 2681 (1991)}.}\footnote[33]{M. Troyer, H. Tsunetsugu, T. M. Rice, J. Riera, and E. Dagotto, "{Spin gap and superconductivity in the one-dimensional $t$-$J$ model with Coulomb repulsion}," \href{https://doi.org/10.1103/PhysRevB.48.4002}{Phys. Rev. B {\bf 48}, 4002 (1993)}.}\footnote[34]{H. Q. Lin, E. Gagliano, and D. K. Campbell, "{Phase Separation in the 1-D Extended Hubbard Model}," \href{https://doi.org/10.1016/S0921-4534(97)01110-6}{Phys. C: Supercond. Appl. {\bf 287}, 1875 (1997)}.}, and also in the systems with strong Hund exchange $J_\mathrm{H}$\footnote[35]{J. Riera, K. Hallberg, and E. Dagotto, "{Phase Diagram of Electronic Models for Transition Metal Oxides in One Dimension}," \href{https://doi.org/10.1103/PhysRevLett.79.713}{Phys. Rev. Lett. {\bf 79}, 713 (1997)}.}.

(b) Fig.~\ref{figS1}\textbf{b}. The nature of the magnetic correlations can be examined with the help of the static structure factor $S(q)$, i.e., from the Fourier transform of the spin-spin correlations $\langle S^z_i S^z_j \rangle$. Fig.~\ref{figS1}{\bf b} depicts the position of the maximum $q_\mathrm{max}$ of $S(q)$. Our results indicate that for $U/W<4$, the correlations have an overall AFM nature ($q_\mathrm{max}\sim\pi$). For $U\to0$, this reflects the paramagnetic state, while for $U\sim W$ AFM and incommensurate-AFM correlations for $x\to0$ and $x\ne0$, respectively. For finite doping $x\ne0$ and for $U/W\gg 1$ (or to be more specific: for $J_\mathrm{H}\gg t$) the system order ferromagnetically ($q_\mathrm{max}=0$) due to the double-exchange mechanism \footnote[30]{C. Zener, "{Interaction Between the $d$ Shells in the Transition Metals}," \href{https://doi.org/10.1103/PhysRev.81.440}{Phys. Rev. {\bf 81}, 440 (1951)}.}\footnote[31]{C. Zener, "{Interaction between the $d$-Shells in the Transition Metals. II. Ferromagnetic Compounds of Manganese with Perovskite Structure}," \href{https://doi.org/10.1103/PhysRev.82.403}{Phys. Rev. {\bf 82}, 403 (1951)}.}. See also Ref.~\cite{Momoi1998,Held2000} of the main text. Note that in our considerations the Hund exchange is given by $J_\mathrm{H}=U/4$, as widely used in many materials such as high-$T_\mathrm{c}$ Fe-based superconductors\footnote[36]{K. Haule and G. Kotliar, "{Coherence–incoherence crossover in the normal state of iron oxypnictides and importance of Hund's rule coupling}," \href{https://doi.org/10.1088/1367-2630/11/2/025021}{New J. Phys. {\bf 11}, 025021 (2009).}}\footnote[37]{Q. Luo, G. Martins, D.-X. Yao, M. Daghofer, R. Yu, A. Moreo, and E. Dagotto, "Neutron and ARPES constraints on the couplings of the multiorbital Hubbard model for the iron pnictides," \href{https://doi.org/10.1103/PhysRevB.82.104508}{Phys. Rev. B {\bf 82}, 104508 (2010).}}\footnote[38]{J. Ferber, K. Foyevtsova, R. Valentí, and H. O. Jeschke, "LDA+DMFT study of the effects of correlation in LiFeAs," \href{https://doi.org/10.1103/PhysRevB.85.094505}{Phys. Rev. B {\bf 85}, 094505 (2012).}}.

\newpage
\section{Charge density oscillations}
\begin{figure*}[!htb]
\includegraphics[width=1.0\textwidth]{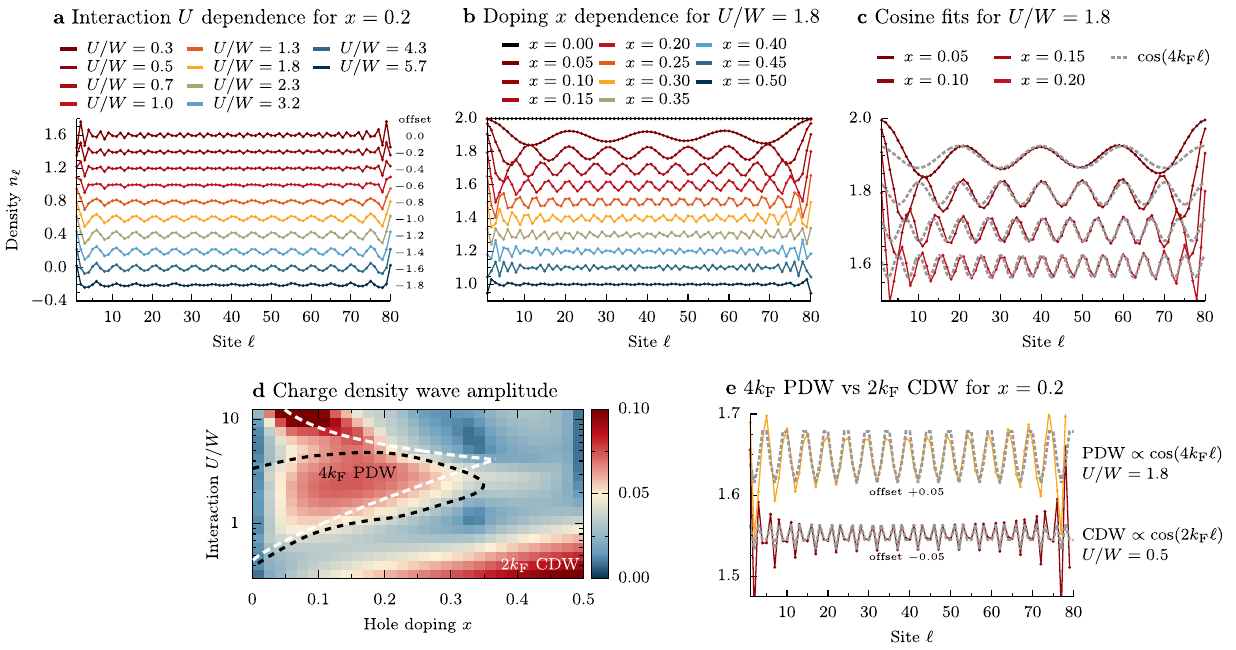}
\caption{{\bf Charge density oscillations.} {\bf a} Interaction $U$ and {\bf b} doping $x$ dependence (for fixed $x=0.2$ and $U/W=1.8$, respectivly) of the spatial profiles of the electron density $n_\ell$. Consecutive curves presented in panel ${\bf a}$ have $-0.2$ offset for clarity. {\bf c} Cosine fits $n_\ell=A\cos(4k_\mathrm{F}\ell)$ with $k_\mathrm{F}=0.5\pi\overline{n}=0.5\pi(1-x)$ to the selected data ($U/W=1.8\,,x=0.05,0.10,0.15,0.20$). {\bf d} Charge density wave amplitude $U$-$x$ phase diagram. {\bf e} Fits to the charge density wave (CDW $2k_\mathrm{F}$ oscillations, $+0.05$ offset for clarity) and pair density wave (PDW $4k_\mathrm{F}$ oscillations, $-0.05$ offset for clarity).}
\label{figS2}
\end{figure*}

In Fig.~\ref{figS2}, we present the analysis of the spatial dependence of the electron density $n_\ell=n_{0\ell}+n_{1\ell}$ at finite doping $x\ne0$. Panels {\bf a} and {\bf b} depict interaction $U$ and doping $x$ dependence, respectively, while panel {\bf c} shows exemplary fits to the $n_\ell=A\cos(4k_\mathrm{F}\ell)$ function.

Interestingly, we find two types of charge density oscillations. At $U\to0$ (in the topologically trivial region), we find "standard" charge density wave (CDW) $2k_\mathrm{F}=\pi\overline{n}=\pi(1-x)$ Friedel oscillations (see fit for $U/W=0.5$ presented in Fig.~\ref{figS2}{\bf e}). On the other hand, in the region where orbital-RVB is stabilized, we find $4k_\mathrm{F}$ oscillations (see fits presented in Fig.~\ref{figS2}{\bf c} and Fig.~\ref{figS2}{\bf e}). The latter phenomenon is consistent with pair-density wave (PDW) with two holes in each minimum in the density (see Ref.~\cite{Berg2010,Zhang2022,Liu2023} of the main text).

\end{document}